\documentclass[aps,prl,twocolumn,showpacs,superscriptaddress,groupedaddress]{revtex4}
\usepackage{txfonts}  
\usepackage{graphicx}  
\usepackage{bm}        
\usepackage{amsfonts}
\usepackage{amssymb}

\begin{document}



\title{Bulk flow scaling for turbulent channel and pipe flows}
\author{Xi Chen}
\affiliation{State Key Laboratory for Turbulence and Complex Systems
and Department of Mechanics, College of
Engineering, Peking University, Beijing 100871, China}
\affiliation{Department of Mechanical Engineering, Texas Tech University, Lubbock, Texas, 79409-1021, USA}
\author{Fazle Hussain}
\affiliation{Department of Mechanical Engineering, Texas Tech University, Lubbock, Texas, 79409-1021, USA}
\author{Zhen-Su She}
\email{she@pku.edu.cn}
\affiliation{State Key Laboratory for Turbulence and Complex Systems
and Department of Mechanics, College of
Engineering, Peking University, Beijing 100871, China}
\date{\today}

\begin{abstract}
We report a theory deriving bulk flow scaling for
canonical wall-bounded flows. The theory accounts for the symmetries of boundary geometry (flat plate channel versus
circular pipe) by a variational
calculation for a large-scale energy length, which characterizes its bulk flow scaling by
a simple exponent, i.e. $m=4$ for channel and 5 for pipe.
The predicted mean velocity shows excellent agreement with several dozen sets of quality empirical data
for a wide range of the Reynolds number (Re), with a universal bulk flow constant
$\kappa\approx0.45$. Predictions for dissipation and turbulent transport in the bulk flow are also given, awaiting data verification.
\end{abstract}

 \maketitle

An intriguing feature of practical turbulent flows is the presence of net spatial momentum and energy transports,
which are constrained by boundaries \citep{Voyage,SmitsMarusic2013}. Flows at the surface of a vessel or inside
a pipeline, over the wings of an aircraft or close to the ground on windfarms, etc., develop abundant flow structures
contributing to turbulent transports. However, even for canonical cases, i.e. flat-plate channels
and cylinder pipes, the boundary effect has not been treated theoretically in predicting mean flow scaling.
A milestone for the mean momentum scaling is the von Karman's velocity-defect law \citep{Ka30}, i.e.
\begin{equation}\label{eq:Defect}
U_c-U=u_\tau g(y/R)
\end{equation}
where $U$ is the streamwise mean velocity; $U_c$ is the mean centerline velocity; $u_\tau$ is the friction velocity
(defined later) and $g$ is an unknown scaling function depending only on the wall distance $y$ (normalized with the
half channel height or pipe radius $R$). Later, Millikan \cite{millikan1938} developed a matching argument deriving the celebrated
Prandtl-Karman log-law of (\ref{eq:Defect}). However, all theoretical accounts have been restricted to the
scaling in an overlap region (typically for $y\le 0.15$), without addressing the influence of geometries on more than 80\% of the flow domain. Most efforts devoted for the whole domain description are empirical, such
as Coles wake function \citep{Coles1956}, composite pade approximation \citep{monkewitz2007,nagib2008}, with
limited accuracy and unknown domain of applicability. A recent work by L'vov et al \cite{Lvov2008} achieves a noteworthy description of channel and pipe flows, but its outer
flow description invoking a fitting function derived from simulation data does not
distinguish channel and pipe. Thus, a theoretical derivation for the complete expression of function $g$
is still missing, which is particularly important to resolve recent vivid debates on the universality of the mean
velocity scaling in the canonical wall-bounded turbulent flows \citep{Marusic2010}.

In this paper, we present a novel attempt which identifies a universal mechanism deriving the mean velocity
scaling for canonical wall-bounded flows, based on a symmetry consideration of wall constraints.
Most importantly, the theory suggests a universal bulk flow constant for channel and pipe.
This is accomplished by introducing a length function whose calculation based on a variational argument yields a geometry dependence (planar versus circular) with an integer scaling exponent (4 for a flat channel and 5 for a cylindrical pipe). The analysis enables a prediction of (\ref{eq:Defect}) valid in the entire flow domain,
and the predicted mean velocity profiles are in excellent agreement with several dozen of recent, reliable experiments over a wide range of $Re$'s. The results also shed
light on the debate between the log-law and power-law \citep{Barenblatt,George06} in favor of the former, and have applications to other boundary effects
(such as roughness, compressibility, pressure gradient - studied by us but not addressed here).

We start with the fully developed incompressible turbulent channel flow, between two parallel plates
of height $2R$, driven by a constant pressure gradient $f_p=-\frac{1}{\rho}{\frac{\partial p}{\partial x}}$ in the
stream-wise x-direction. The flow develops a mean-velocity profile $U$, depending on wall-normal y-direction
only. The mean momentum flux is described by the Reynolds averaged Navier-Stokes (RANS) equation, i.e.
\begin{equation}\label{eq:MMED}
\nu{\partial_y U}-\overline{u'v'}=\tau_p.
\end{equation}
where $S\equiv \partial_y U$ is mean shear; $W\equiv -\overline{u'v'}$ denotes Reynolds stress which is unknown;
$\tau_p=\int^r_0 f_p dr'=u^2_\tau r/R$ is the total stress with $r=R-y$, the distance to the centerline, and
$u_\tau\equiv \sqrt{f_p R}$ is the friction velocity determined by the pressure force. Note that
dimensionally, (\ref{eq:MMED}) has an alternative interpretation at the local position $r$: the pressure gradient
force supplies the energy $\tau_p=f_pr$ which balances the viscous damp $\nu S$ and the turbulent shear fluctuation $W$. In this sense,
for a cubic flow volume in a channel, namely $V_r=r \times R\times R $ ($R\times R$ indicates the $r$-surface area),
the total turbulent shear fluctuation energy is thus $M=\int_0^{V_r} W d {\textbf{v}}$
(we will return to this quantity later).

Then, the product of viscous and Reynolds stresses contributes to the growth of
the turbulent kinetic energy $k=\overline{u'_iu'_i}/2$, which is described by the mean kinetic energy equation, i.e.
\begin{equation}\label{eq:MKED}
        {S}{W} + {\Pi } = {\varepsilon}.
\end{equation}
Here $\mathcal{P}=SW$ is the production; $\Pi$ represents the spatial energy transfer (including diffusion, convection and
fluctuation transport); $\varepsilon$ the viscous dissipation (for explicit expressions, see \citep{Davidson2004}).

In our recent work \cite{Chen2016}, a dimensional analysis among $\varepsilon, S, W$ yields
\begin{equation}\label{eq:elln}
\ell= W^{(\frac{1}{n}+\frac{1}{2})} S^{(\frac{1}{n}-1)} \varepsilon^{(-\frac{1}{n})},
\end{equation}
where $\ell$ is the characteristic length representing eddies responsible for the energy spatial transfer, and $n$ is an arbitrary real number, tentatively chosen to be
integer. Note that as $n \rightarrow \infty$, the length becomes the classical
mixing length of Prandtl: $\ell_\infty=\sqrt{W}/S$; while a unique $n=4$ defines a physically
meaningful length valid throughout the channel, i.e.
\begin{equation}\label{eq:elleps}
 \ell_\varepsilon= {W^{\frac{3}{4}}}{S^{\frac{-3}{4}}}\varepsilon^{\frac{-1}{4}} .
\end{equation}
This length is similar to the crucial scaling
function in the model of L'vov et al. \citep{Lvov2008} (restricted to the bulk zone only), but its interpretation follows a concept of
order function developed by us \citep{She2010}. Our main result of this paper is to give a physical derivation of $\ell_\varepsilon$.

\begin{figure}
\centering
\includegraphics[width = 5 cm]{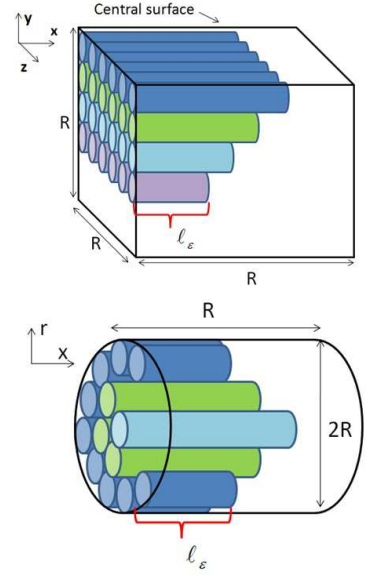}   
    \caption{Sketch of PE (tubes) with streamwise length scale $\ell_\varepsilon$ (depending on centerline distance) in a channel flow volume $V_R=R^3$ (left), and a pipe flow volume $V_R=\pi R^3$ (right).}
    \label{fig:CHPP}
 \end{figure}

According to its definition (\ref{eq:elleps}), $\ell_\varepsilon$ can also be expressed in terms of the
eddy viscosity $\nu_t=W/S$, i.e. $\ell_\varepsilon=\nu^{3/4}_t/\varepsilon^{1/4}$. This expression reminds us of
the Kolmogorov dissipation length
$\eta=\nu^{3/4}/\varepsilon^{1/4}$, where $\nu$ is the kinematic viscosity. Following the interpretation of $\eta$,
$\ell_\varepsilon$ is presumably related to turbulence production eddies (since $SW=P$), hence called the length of production eddies (PE). In analogy to Townsend \cite{Townsend}, a possible materialization of PE is through the ensemble averaged vortex packets \cite{Adrian} or clusters \cite{Jimenez}. Moreover, recalling the attached eddy hypothesis by Perry \cite{Perry1982}, we assume similarly that PE distribute uniformly in the spanwise direction (since time average is applied), and the streamwise characteristic size is $\ell_\varepsilon$ - only depending on centerline distance $r$. In figure \ref{fig:CHPP}, we depict the distribution of PE in a channel (and a pipe), where one can see a monotonic decrease of $\ell_\varepsilon$ away from centerline, indicating the influence of wall constraint $\ell_\varepsilon=0$ approaching the wall. Note that one can also interpret the increment of $\ell_\varepsilon$ approaching the centerline as the growth of PE, in analogy to the growth of the wall attached eddies as wall distance increases \cite{Perry1982}. Under such a statistical point of view, we will consider two important quantities associated with the PE, which are the total shear fluctuation energy $M'$ (from Reynolds shear stress), and the total kinetic energy $E'$ associated with the growth of PE, as presented below.

Recall that the total shear fluctuation energy for a given flow volume $V_r$ is $M$. Then, for the shear fluctuation energy
of PE sketched in figure \ref{fig:CHPP}, an volume integration of $W$ yields
$M'=\int_0^{r} W \ell_\epsilon R dr=\int_0^{V_r} W (\ell_\epsilon/R) d\textbf{v}$. Moreover, as $\nabla\ell_\varepsilon$
represents the growth rate of PE size as $y$ increases, and $u_\tau$ is a global velocity scale, thus
$u_\tau \nabla\ell_\varepsilon$ represents the increment momentum due to the PE's growing size, and
$E'=\int_0^{V_r} | u_\tau \nabla\ell_\varepsilon|^2 d {\textbf{v}}$ is the total kinetic energy associated with the growth of PE.

Now, a variational argument is postulated by assuming that for a given $M'$ (determined by the pressure force since
$W\approx \tau_p$), $E'$ should be minimum - as the flow reaches a quasi-equilibrium state. In other words, turbulent
fluctuations dissipate kinetic energy, resulting in a minimum of $E'$ for the growth of PE. As both $M'$ and $E'$ depend
on $\ell_\varepsilon$, following the calculus of variations \citep{Francis1976}, we thus require for all infinitesimal
variations $\delta\ell_\varepsilon$,
\begin{eqnarray}\label{eq:Var}
\delta E'-\alpha \delta M'=0,    
\end{eqnarray}
where $\alpha$ is a dimensionless
Lagrange multiplier. Generally, $\alpha$ depends on $V_r$ since $M'$ and $E'$ are
integrated over $V_r$. To nondimensionalize $V_r$, a simple choice is $\alpha=\alpha_0 {V_r}/{V_R}$,
where $\alpha_0$ is a constant and $V_R=R^3$ is the total cubic volume. It turns out that such a constant $\alpha_0$ assumption
is supported by the results shown later.

\begin{figure}
\centering
\includegraphics[width = 8 cm]{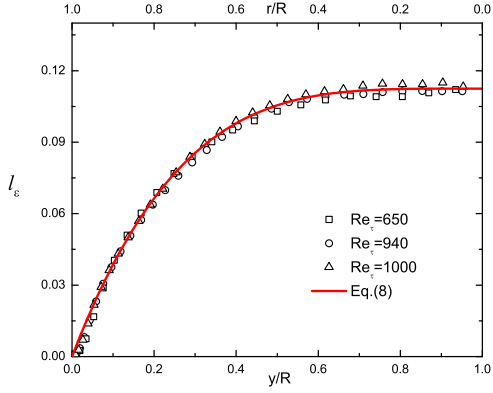}
    \caption{Characteristic length $\ell_\epsilon=\ell_\infty/\Theta^{1/4}$ at different $Re$'s in channels. Bottom axis is the wall distance $y/R$ and the up axis is the centerline distance $r/R$. The curve (\ref{eq:CHlnu}) exhibits convincing data collapse with $\kappa=0.45$. Symbols are DNS data of $\ell^{DNS}_\infty$ divided by theoretical $\Theta$ with $r_c=0.27$. Dada are from Iwamoto \emph{et al} \citep{Iwamoto2002} ($Re_\tau=650$); Hoyas \& Jimenez \citep{Hoyas2006} ($Re_\tau=940$); Lee \& Moser \cite{Moser2015}($Re_\tau=1000$).}
    \label{fig:Ell}
 \end{figure}

By substituting $\delta E'=-2\int_0^{V_r} u_\tau^2\Delta\ell_\varepsilon\delta\ell_\varepsilon d{\textbf{v}}$ and
$\delta M'=\int_0^{V_r}(W/R)\delta\ell_\varepsilon  d{\textbf{v}}$ into (\ref{eq:Var}), we thus obtain a diffusion
equation for $\ell_\varepsilon$ (eliminating the volume integral),
\begin{equation}\label{eq:Var2}
-2u_\tau^2\Delta\ell_\varepsilon-\alpha W/R=0.
\end{equation}
With $W\approx\tau_p=u^2_\tau r/R$, $\alpha=\alpha_0 V_r/R^3=\alpha_0 r/R$, we have $2\Delta\ell_\varepsilon\approx-\alpha_0 r^2/R^3$,
which, after integrating with $r$, leads to $\ell_\varepsilon/R\approx-\alpha_0 (r/R)^4/24+a_1 r/R+a_2$. Note that $a_1=0$
due to the central symmetry ($\partial_r\ell_\varepsilon=0$ at $r=0$), and $a_2=\alpha_0 /24$ due to the wall condition
($\ell_\varepsilon=0$ at $r=R$). Therefore, $\ell_\varepsilon$ in a channel flow is:
\begin{equation}\label{eq:CHlnu}
\ell^{CH}_\varepsilon/R\approx{\kappa}(1-r'^4)/4
\end{equation}
where $\kappa=\alpha_0/6$ and $r'=r/R$ is substituted. (\ref{eq:CHlnu}) is validated in figure \ref{fig:Ell} using direct numerical
simulation (DNS) data. It is clear that different $Re$'s
profiles of $\ell_\varepsilon$ collapse and agree well with (\ref{eq:CHlnu}) almost in the entire flow domain.
The only empirical parameter $\kappa\approx0.45$ is extracted from DNS data, indicating $\alpha_0=6\kappa\approx 2.7$,
a constant expected by the preceding analysis.

All above analysis can be equally applied to turbulent pipe flow. The difference from channel is that, the
flow volume corresponding to the cylindrical boundary (see figure \ref{fig:CHPP}) is $V_r=R \pi r^2 $, and $M' = \int_0^r {{W}} {\ell _\varepsilon }2\pi rdr = \int_0^{{V_r}} {{W}} ({\ell _\varepsilon }/R)d\textbf{v}$, $V_R=\pi R^3$.
In this case,  $\alpha=\alpha_0 r^2/R^2$, and a similar calculation of (\ref{eq:Var2}) for pipe flow yields
\begin{equation}\label{eq:Pipelnu}
\ell^{Pipe}_\varepsilon/R\approx{\kappa}(1-r'^5)/5.
\end{equation}
Here the coefficient $\kappa/5$ is determined by requiring the same near wall asymptotic scaling as (\ref{eq:CHlnu}),
i.e. $\ell^{Pipe}_\varepsilon/R\approx\ell^{CH}_\varepsilon/R\approx \kappa (y/R)$ when $r'\rightarrow1$. This is reasonable
because the geometry effect vanishes close to the wall, hence channel and pipe should share the same $\kappa$ (validated later). Also note that for turbulent boundary layer (TBL), though $W$ may different from channel to TBL, a leading order expansion $W\propto r/R$ ($r$ is the distance to the boundary layer thickness $R$) in TBL would also lead to (\ref{eq:CHlnu}) based on (\ref{eq:Var2}). In other words, (\ref{eq:CHlnu}) in channel should also apply in TBL when the latter flow becomes nearly parallel (at large $Re$'s); this result is in consistent with the same flat plate wall condition in the two flows.

\begin{figure}
\centering
\includegraphics[width = 9 cm]{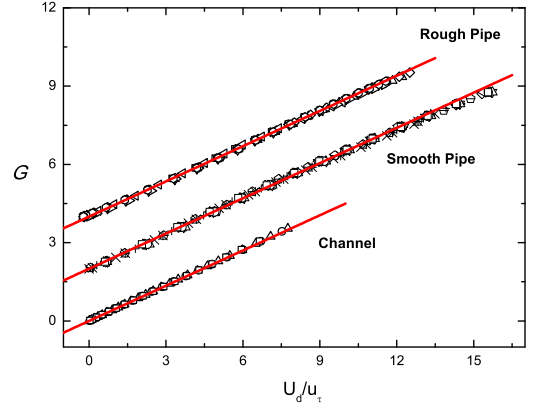}
    \caption{Empirical $U_d/u_\tau$ versus theoretical scaling function $G$ for 25 profiles in the bulk flow region
    ($50\leqslant yu_\tau/\nu\leqslant Re_\tau$), elucidating a good linear relation with a universal slope $\kappa=0.45$
    (lines) in (\ref{eq:Ud2}) for $Re_\tau$ varying over three decades. Profiles are vertically staggered for clarity.
    There are 3 profiles from DNS channel (data are the same as in figure \ref{fig:Ell}); 14 profiles from Princeton
    smooth pipe with $Re_\tau$ from $6\times10^3$ to $5\times10^5$ \citep{Mckeon2004} and 8 from rough pipe with $Re_\tau$
    from $1\times10^4$ to $2\times10^5$ \citep{Shockling2006}.}
    \label{fig:Ud}
 \end{figure}

Interestingly, a joint solution of (\ref{eq:MMED}) and (\ref{eq:MKED}) can be obtained using $\ell_\varepsilon$ and
$\Theta=\varepsilon/(SW)$. From (\ref{eq:elleps}) one has $S=\sqrt{W}/(\ell_\varepsilon\Theta^{1/4})$.
Integrating $S$ with $r$ and using $W\approx{\tau_p}$, one obtains the velocity-defect
law (\ref{eq:Defect}), i.e.
\begin{equation}\label{eq:Ud}
U_c-U=\int_0^r Sd\hat{r}\approx \int_0^r \frac{\sqrt{\tau_p}}{\ell_\varepsilon\Theta^{1/4}} d\hat{r}
\end{equation}
Here $\Theta=[{1 + {({r_{c}}/r')^2}}]/{(1 + r_{c}^2)}$ has been derived in \cite{Chen2016}; it connects two asymptotes,
i.e. $\Theta\rightarrow1$ as $r'\rightarrow1$ and $\Theta\rightarrow 1/r'^2$ as $r'\rightarrow0$ smoothly, valid for the entire
flow domain. Note that (\ref{eq:Ud}) also rewrites as
\begin{eqnarray}\label{eq:Ud2}
\kappa U_d/u_\tau=G(r')\approx\int_0^{r'} {f(\hat{r})} d\hat{r},
\end{eqnarray}
where $U_d=U_c-U$, $f={{m{{[(1 + {r_c}^2)/(\hat{r}{^2}+{r_c}^2)]}^{1/4}}{\hat{r}} }}/{{(1 - \hat{r}{^m})}}$ ($m=4$
for channel and 5 for pipe). The parameter $r_c$ indicates the thickness of the core layer in channel and pipe
(zero in TBL due to the absence of opposite wall), which has a slight $Re$-dependence at moderate $Re$'s. It is obtained by
fitting $G$ with the velocity-defect data, which yields $r_c\approx0.27$ for channels (DNS) and $r_c\approx0.5$ for Princeton
pipes (EXP). The predicted velocity defect is shown in figure \ref{fig:Ud}, where the
universal bulk flow constant $\kappa\approx0.45$ is remarked by the linear slope agreeing well with 25 sets of
mean velocity profiles. Note that the result for smooth pipe is also applied to rough pipe (with the same $\kappa$ and $r_c$),
consistent with the Townsend's similarity hypothesis \citep{Townsend}.

\begin{figure}
\centering
\includegraphics[width = 9 cm]{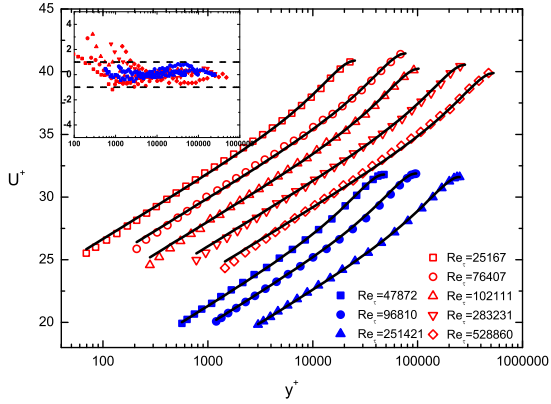}
    \caption{Mean velocity data (symbols) compared with (\ref{eq:Ud2}) (lines), using empirical $U_c/u_\tau$, $\kappa$ and
    $r_c$. Open symbols indicate smooth pipes \citep{Mckeon2004} while solids indicate rough pipes \citep{Shockling2006}.
    Inset shows relative errors (times 100) which are mostly bounded within 1\% (dashed lines). Profiles for smooth pipes are
    vertically staggered for clarity.}
    \label{fig:MVP}
 \end{figure}

It is important to note that our current results support the the asymptotic log-law instead of power-law. Note that at large $Re$'s, there would be an
asymptotic interval where $\Theta\approx1$, $\tau_p\approx u^2_\tau$ and $\ell_\varepsilon\approx\kappa y$. In this
region, $\ell_\varepsilon\approx\ell_\infty$, so the present calculation is identical to the Prandtl-Karman's log-law \citep{Voyage,Davidson2004}, $\ell_\infty\approx \kappa_{K} y$.
This implies that the bulk flow constant $\kappa$ derived in the present study is exactly the Karman constant $\kappa_{K}$. Furthermore, rewriting (\ref{eq:Ud2}) as $U/u_\tau=U_c/u_\tau-G/\kappa$, and the leading order contribution in $G$ is $\int^{1-y/R}_0 1/(1-\hat{r}) d\hat{r}\propto \ln(y/R)$, we thus have $U/u_\tau-U_c/u_\tau\propto \kappa^{-1}\ln (y/R)$, indicating the logarithmic scaling. Thus our results support
the asymptotic log-law. In figure \ref{fig:MVP}, using the empirical
$U_c$ (and $\kappa$, $r_c$), we plot the mean velocity by (\ref{eq:Ud2}), displaying impressive agreement with data (smooth and rough pipes only; channel data also agree well, but not shown). The relative errors are bounded within $1\%$ (except for first several points near wall) - the same level as the data
uncertainty. Note that the inner bound of (\ref{eq:Ud2}) reaches as close to the wall as $y u_\tau/\nu\approx50$ (buffer layer thickness). The correction within the buffer layer needs wall function, to be reported elsewhere.

Finally, the dissipation and transport in the mean kinetic energy equation (\ref{eq:MKED}) are predicted:
\begin{eqnarray}
\varepsilon  = SW\Theta  \approx \frac{u_\tau^3}{R} \frac{{m{{(r{'^2} + r_c^2)}^{3/4}}}}{{\kappa (1 - r{'^m}){{(1 + r_c^2)}^{3/4}}}}\nonumber\\
\Pi {\rm{ = }}SW(\Theta  - 1) \approx \frac{u_\tau^3}{R} \frac{{mr_c^2(1 - r{'^2}){{(r{'^2} + r_c^2)}^{ - 1/4}}}}{{\kappa (1 - r{'^m}){{(1 + r_c^2)}^{3/4}}}}
\end{eqnarray}
Particularly, the
centerline dissipation (equaling centerline transport) is:
\begin{equation}
{\varepsilon _0}=\Pi_0\approx \frac{u_\tau^3}{R} \frac{m r_c^{3/2}}{\kappa {(1 + r_c^2)}^{3/4}}
\end{equation}
With the value $\kappa\approx0.45$ and $r_c\approx0.5$, we have ${\varepsilon _0} \approx 3.3u_\tau^3/R$ for pipe.
For current channel data, using $\kappa\approx0.45$ and $r_c\approx0.27$ yields ${\varepsilon _0}\approx1.2 u_\tau^3/R$.
However, since $r_c$ has a moderate $Re$-effect (indicating a growth of central core layer), the predicted centerline dissipation
may increase with increasing $Re$. Assuming the same $r_c$ for high $Re$ channels and pipes, one would have
${\varepsilon^{Pipe} _0}/{\varepsilon^{CH} _0}=5/4$. These await verifications when ${\varepsilon _0}$ data are available.

In summary, we have developed an analytical theory for joint closures of the mean momentum and kinetic energy equations for
turbulent channel and pipe flows. The variational assumption leads to an analytical formula of the eddy length function,
where a universal bulk flow constant $\kappa\approx0.45$ is identified to be valid for the
entire flow domain (much beyond the overlap region). Note that (\ref{eq:CHlnu}) and (\ref{eq:Pipelnu}) indicate a breaking
of dilation invariance for $\ell_\varepsilon$ because of the presence of a characteristic length $\ell_0=\kappa R/m$ at the
centreline. However, the dilation invariance is preserved in its gradient, i.e. $d\ell _\varepsilon /dr' \propto r{'^{m - 1}}$.
Such a symmetry perspective is further explored in connection of the
dilation symmetry in the direction normal to the wall for turbulence model equations widely used in engineering applications (i.e.
$k-\omega$ equation), which will be reported elsewhere.

This work is supported by National Nature
Science (China) Fund 11452002 and 11521091 and by MOST (China) 973 project 2009CB724100.

\appendix

\end{document}